\documentclass[preprint]{emulateapj}
\usepackage{amsmath}
\usepackage{array}
\usepackage{float}
\usepackage{graphicx}
\usepackage{subfigure}
\usepackage{natbib}
\usepackage{color}

\begin{document}
\title{The bimodal color distribution of small Kuiper Belt objects\footnotemark[*]}
\author{Ian Wong and Michael E. Brown}
\affil{Division of Geological and Planetary Sciences, California Institute of Technology,
Pasadena, CA 91125, USA; iwong@caltech.edu}

\footnotetext[*]{Based on data collected at Subaru Telescope, which is operated by the National Astronomical Observatory of Japan.}

\begin{abstract}
We conducted a two-night photometric survey of small Kuiper Belt objects (KBOs) near opposition using the wide-field Hyper Suprime-Cam instrument on the 8.2~m Subaru Telescope. The survey covered about 90~deg$^2$ of sky, with each field imaged in the $g$ and $i$ bands. We detected 356 KBOs, ranging in absolute magnitude from 6.5 to 10.4. Filtering for high-inclination objects within the hot KBO population, we show that the $g-i$ color distribution is strongly bimodal, indicative of two color classes --- the red and very red subpopulations. After categorizing objects into the two subpopulations by color, we present the first dedicated analysis of the magnitude distributions of the individual color subpopulations and demonstrate that the two distributions are roughly identical in shape throughout the entire size range covered by our survey. Comparing the color distribution of small hot KBOs with that of Centaurs, we find that they have similar bimodal shapes, thereby providing strong confirmation of previous explanations for the attested bimodality of Centaurs. We also show that the magnitude distributions of the two KBO color subpopulations and the two color subpopulations observed in the Jupiter Trojans are statistically indistinguishable. Finally, we discuss a hypothesis describing the origin of the KBO color bimodality based on our survey results.

\end{abstract}
\emph{Keywords:} Kuiper Belt: general, minor planets, asteroids: general

\section{Introduction}

Obtaining a better understanding of the composition of Kuiper Belt objects (KBOs) has become a major field of interest in planetary science, with important implications for formation and evolution theories of the solar system. Over the past decade, a concerted observational effort has produced a significant increase in the number of known objects and revealed the dynamical complexity of the Kuiper Belt region. However, even with recent advances in telescope capabilities, the great heliocentric distances of these bodies continue to be an obstacle for detailed characterization of their surfaces. At present, only the largest KBOs are amenable to spectroscopic study with sufficient signal-to-noise to distinguish absorption features of chemical species. 

While spectroscopic observations of large KBOs have revealed a wide range of absorbing species \citep[such as ammonia, methanol, methane, and ethane; see, for example, review by][]{brownreview}, the surface compositions of these bodies are not expected to reflect directly on the primordial makeup of the protoplanetary disk in the trans-Neptunian region. These large objects, with diameters exceeding $\sim$500~km, have sufficient mass to be gravitationally circularized and are likely internally differentiated. As a result, their surfaces have been shaped by secondary processes (including the possible retention of tenuous and/or ephemeral atmospheres) and may differ significantly from their bulk composition. To obtain a more representative picture of the bulk composition of KBOs, we must study smaller objects, whose surfaces are more representative of the primordial planetesimals in the trans-Neptunian region.

Broadband photometry has served as a powerful tool for understanding the surface properties of smaller KBOs. An analysis of the range of colors within the overall KBO population and trends in the color distribution with relation to the various dynamical classes within the Kuiper belt region provides a descriptive picture of the compositional diversity of the formation environment and how the Kuiper Belt evolved to its current state. To date, hundreds of KBO colors have been measured, revealing the general properties of low albedos and reddish colors across the intermediate and small size ranges, extending down through objects with diameters of 50~km and less \citep{hainaut}. Several studies have suggested a color bimodality among objects in the dynamically-excited (i.e., hot) KBO population \citep[e.g.,][]{fraserbrown,peixinho,lacerda}. A bimodal color distribution would have major implications for the formation of these bodies, indicating two types of surface compositions and perhaps two different formation environments. However, since most of the color measurements in the literature have been obtained through targeted observations, the observational biases are difficult to quantify, and as such, the underlying color distribution of hot KBOs remains poorly constrained.

Photometric observations of Centaurs, which are former members of the hot KBO population  that have been scattered onto giant planet crossing orbits, provide a complementary look at the color distribution of these outer solar system objects. Their smaller heliocentric distances allow greater ease in obtaining precise color measurements. Recent analyses of Centaurs demonstrate a robust bimodality in the optical color distribution extending through the entire observed size range --- from the largest Centaur, 10199 Chariklo ($\sim$250~km), down through $\sim$$10-20$~km sized objects --- dividing the Centaur population into two color classes \citep[e.g.,][]{peixinho2,barucci3,perna}. 

Some have argued that, based on the putative color bimodality of similarly-sized hot KBOs, the two color classes in the Centaurs correspond to the two color classes observed in the Kuiper Belt \citep{fraserbrown,peixinho,peixinho2015}. Others have posited that the bimodality in the Centaur color distribution could have arisen from divergent orbital histories of these objects as they were scattered inward into their present-day eccentric orbits: objects that spent a larger fraction of the time at small heliocentric distances would have experienced on average higher temperatures and space weathering, which may have altered their surfaces \citep{melita}. Without a fuller understanding of the underlying color distribution of KBOs in this size range, the origin of the Centaur color bimodality has remained an open question.

\begin{figure*}[t!]
\begin{center}
\includegraphics[width=17cm]{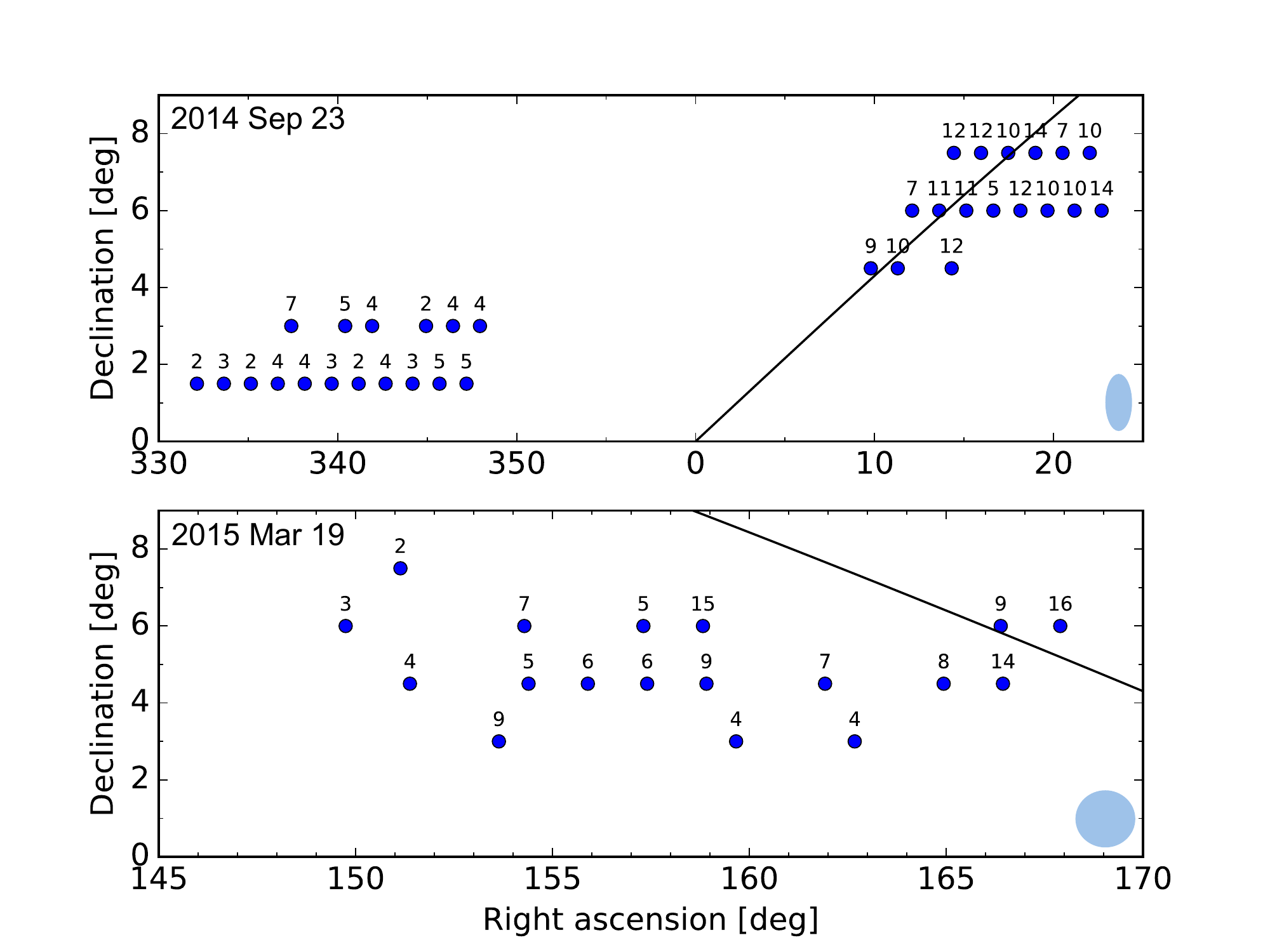}
\end{center}
\caption{Locations of the 52 Subaru Hyper Suprime-Cam fields observed as part of our two-night Kuiper Belt color survey (blue points). The position of the ecliptic on each night is indicated by the solid black line. The number of objects detected in each field is given by the number above the corresponding point. The shaded ellipse shown in the lower right corner of each panel represents the field of view of each exposure, to scale.} \label{positions}
\end{figure*}

We present the results of the first systematic photometric survey of small KBOs, conducted using the Hyper Suprime-Cam instrument at the Subaru Telescope.This instrument has the largest field of view of any camera on an 8--10~m class telescope currently in operation, making it the ideal tool for surveying small Kuiper Belt objects. By imaging fields in two bandpasses, we were able to simultaneously detect KBOs (on average 7 per field) and measure their colors to efficiently achieve a uniform survey dataset. Filtering our dataset to study the hot KBO population specifically, we report a robust bimodality in the color distribution. The paper is organized as follows: The observing strategy and data processing methods of our survey are described in Section~\ref{sec:obs}. In Section~\ref{sec:analysis}, we analyze the color distribution of hot KBOs and categorize objects into the two color classes in order to study their magnitude distributions individually. Finally, in Section~\ref{sec:discussion}, we discuss the survey results in the context of previous analyses of Centaurs and Jupiter Trojans and present a hypothesis of the origin of the color bimodality.

\section{Observations}\label{sec:obs}

We carried out observations of the Kuiper Belt near opposition using the Hyper Suprime-Cam instrument (HSC) on the 8.2~m Subaru Telescope at Mauna Kea, Hawaii during the nights of UT 2014 September 23 and 2015 March 19. The HSC is a wide-field prime focus camera that consists of a mosaic of 104 2048$\times4096$~pixel CCD detectors and covers a $90'$ diameter field of view with a pixel scale of $0.17''$ \citep{miyazaki}. We obtained four 180~s exposures in each of our observing fields --- twice in the $g$ filter ($\lambda_{\mathrm{eff}}=480.9$~nm) and twice in the $i$ filter ($\lambda_{\mathrm{eff}}=770.9$~nm). Fields were imaged in groups following the band sequence $i-i-g-g$ or $g-g-i-i$ in order to minimize the number of filter changes. The average observational arc for each field is 230~minutes, with the four exposures in each field spaced roughly evenly at 1~hour intervals. The time interval between exposures did not vary significantly from field to field or between the two nights of observation. Figure~\ref{positions} shows the on-sky positions of the 52 observed fields, with a total surveyed area of 92~deg$^2$. Full details of our observations and computed orbital, magnitude, and color information are provided in a supplemental table.

We processed the images and detected moving objects using methods similar to those described in detail in our previous Subaru/Suprime-Cam survey of small Jupiter Trojans \citep{wong2}. After bias-subtracting and flat-fielding the images, we calculated the astrometric solution for each chip by first creating a pixel position catalog of all the bright sources using Version 2.11 of SExtractor \citep{sextractor} and then feeding the catalog into Version 1.7.0 of SCAMP \citep{scamp} to compute the astrometric projection parameters. The typical residual RMS value between the on-sky positions of detected stars in an image and matched reference stars is much less than $0.5''$; the typical RMS of the position offset between matched stars across images taken in the same observing field is less than $0.05''$.

After passing the distortion-corrected images through SExtractor again, with the detection threshold set at 1.5 times the background standard deviation, we systematically searched for transients, i.e., sets of four source detections (one in each of the four images in an observing field) that do not reoccur in the same location. We carried out orbital fits for all transients using the methods described in \citet{bernstein}. A transient was classified as a moving object candidate if the corresponding $\chi^{2}$ value from the orbit fit was less than 15. We considered only transients consistent with apparent sky motion less than $10''/\mathrm{hr}$; this cutoff was chosen with the intent of detecting some closer-in Centaurs, in addition to KBOs.

\begin{figure}[t!]
\begin{center}
\includegraphics[width=9cm]{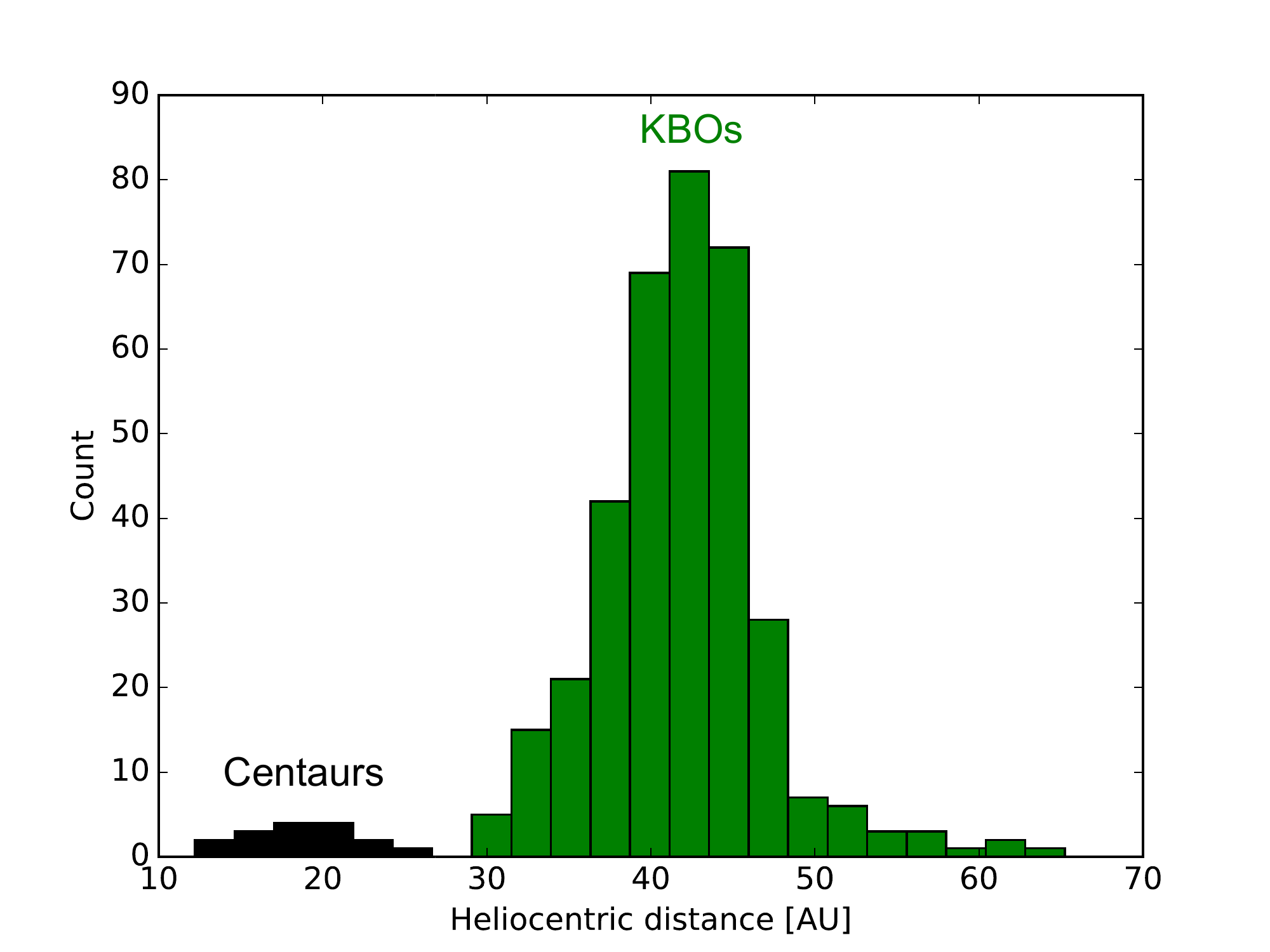}
\end{center}
\caption{Distribution of heliocentric distance for the 372 asteroids detected in our survey. Objects with heliocentric distances greater than 30~au are classified as KBOs, while the closer-in objects are Centaurs. The typical uncertainty of our heliocentric distance estimates is 3.5~au.}\label{heliodist}
\end{figure}

We visually verified every moving object candidate, rejecting all non-asteroidal moving object candidates, as well as asteroids that passed through regions of the chip that would likely yield unreliable and uncorrectable magnitude measurements (e.g., diffraction spikes, chip artifacts). Our survey detected a total of 372 validated objects. Of these objects, 18 passed in front of background stars, coincided with a cosmic ray hit, or passed close to the edge of a chip in one out of the four exposures. For these objects, we removed the problematic detections from consideration in computing heliocentric distances and measuring magnitudes. The distribution of heliocentric distances (computed in our orbital fits) for all 372 objects is shown in Figure~\ref{heliodist}. The median uncertainty in the calculated heliocentric distances is 3.5~au.

We considered all 356 objects with heliocentric distances larger than the orbital semi-major axis of Neptune (roughly 30~au) to be KBOs, with the remaining 16 closer-in objects identified as Centaurs. The separation between KBOs and Centaurs is evident in the plot of heliocentric distances. In our analysis, we also sought to filter out possible contamination from cold classical KBOs, which are characterized by photometric properties, orbital properties, and a size distribution that are distinct from the rest of the Kuiper Belt \citep[see, for example, the review by][and references therein]{brownreview}. 

Since cold classical KBOs have low orbital inclinations, we wanted to remove from consideration all objects with inclinations less than $5^{\circ}$ \citep[as was done in, for example,][]{bernstein2}. Due to our survey's short observational arcs, the uncertainties in calculated inclinations are quite large, averaging around $10^{\circ}$. To select for hot KBOs, we filter for objects observed at ecliptic latitudes greater than $5^{\circ}$ and/or objects with measured inclinations greater than $5^{\circ}$ at the $1\sigma$ level (i.e., $i-\Delta i > 5^{\circ}$). The resulting final set of hot KBOs contains 136 objects and serves as the operational KBO dataset for our analysis. In the following, we will refer to this dataset as simply the survey KBOs.

We measured the flux of each object through a fixed circular aperture with a diameter of 5~pixels, which roughly corresponds to the average seeing on the nights of our observations. The apparent magnitude $m$ of an object in an image is $m = m_{0}-2.5\log_{10}(f) = m_{0}+m_{s}$, where $f$ is the measured flux, $m_{0}$ is the zero-point magnitude of the image, and the survey magnitude has been defined as $m_{s}\equiv -2.5\log_{10}(f)$. For each exposure, following the methods of \citet{wong2}, we calculated the zero-point magnitude of each chip by matching all non-saturated sources detected by SExtractor to reference stars in the Sloan Digital Sky Survey Data Release 9 (SDSS DR9) catalog and carrying out a linear fit with slope set to unity through the points $(m_{s},m)$. When calibrating images in the HSC $g$ and $i$ filters, we used SDSS \textit{g}-band and \textit{i}-band star magnitudes, respectively. The resulting apparent magnitudes are therefore benchmarked to the SDSS photometric system. We computed the corresponding uncertainties in the apparent magnitude using standard error propagation techniques.

We compared the positions of the KBOs detected by our survey with those of previously discovered KBOs listed in the Minor Planet Center catalog for the nights of our observations. Ten objects were matched to known KBOs (see Supplemental Table).

\section{Data analysis}\label{sec:analysis}

\subsection{KBO color distribution}\label{subsec:colors}

From the apparent \textit{g} and \textit{i} magnitudes obtained through our photometric calibration (generally two in each band, unless masked due to a problematic detection in one exposure), we computed error-weighted mean \textit{g} and \textit{i} magnitudes for each object, from which we subsequently derived the $g-i$ color. The uncertainty in color $\sigma_{g-i}$ is given by the quadrature sum of the individual \textit{g} and \textit{i} mean magnitude errors.

Asteroid rotation during the approximately 4~hour observational arc of each detected object can entail an additional contribution to the color uncertainty. We followed the methods of \citet{wong2} and utilized the repeated observations in each band to empirically determine the rotational contribution to the color uncertainty. The standard deviation of differences between two consecutive $g$-band magnitude measurements is comprised of a contribution from the photometric uncertainties of the individual magnitude measurements alone (given by the quadrature sum of the individual magnitude uncertainties), as well as any additional contribution from magnitude modulation of the object due to rotation. When comparing the standard deviation of $g$-band magnitude differences (binned by 0.5~mag intervals) with the corresponding photometric error contribution, we found that the values were comparable; in other words, the photometric error contribution does not systematically underestimate the standard deviation values. The same conclusion was reached for $i$-band. Therefore, we find that asteroid rotation does not have an appreciable effect on the color measurements overall.

Only two KBOs detected by our survey and matched with known objects have previous color measurements --- 2000 QN251 and 2005 SC278. Both of these KBOs were observed and analyzed by \citet{sheppard2012}, who reported $g-i$ colors of $1.21\pm 0.07$ and $1.51 \pm 0.04$, respectively. The colors that we measured are $1.73\pm 0.05$ and $1.40 \pm 0.03$, respectively. While the colors measured for 2005 SC278 are comparable, our measured color for 2000 QN251 is much redder than the earlier measurement.

2000 QN251 is one of the brightest objects detected by our survey, with calculated $g,g,i,i$ magnitudes of 24.3, 24.4, 22.6, 22.6, respectively. All four images had comparable seeing, and our computed absolute magnitude (derived from interpolation of $g$- and $i$-band magnitudes) matches the value listed by the Minor Planet Center. Our reported magnitude values preclude any rotational modulation in apparent magnitude between consecutive images taken in the same band. We suggest that the discrepancy between the measured colors in our work and \citet{sheppard2012} may be due to inconsistent observing conditions during the previous observation or a different observing cadence, which may have led to a rotational contribution to the \citet{sheppard2012} color measurement. We note that neither 2000 QN251 nor 2005 SC278 is a hot KBO, so they are not included in the color-magnitude analysis in this paper.

\begin{figure}[t!]
\begin{center}
\includegraphics[width=9cm]{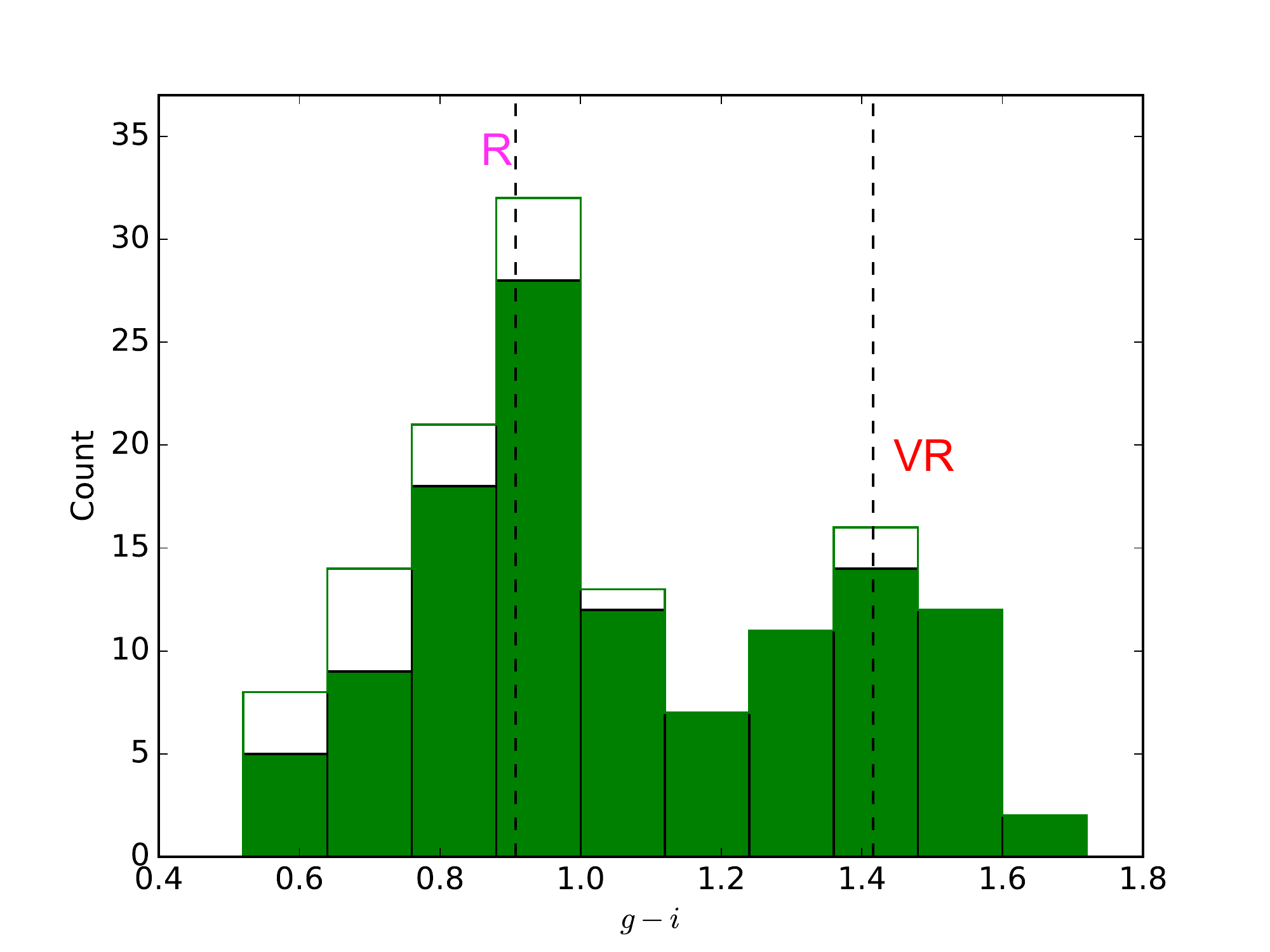}
\end{center}
\caption{Histogram of $g-i$ colors for hot KBOs detected in our survey: the unfilled graph shows the distribution of all 136 objects in the dataset, while the filled graph shows the distribution of only the 118 objects with color uncertainties $\sigma_{g-i} \le 0.15$. A robust bimodality is evident in both distributions, dividing the objects into two subpopulations --- red (R) and very red (VR). The vertical dashed lines indicate the computed mean colors of the R and VR subpopulations --- $\overline{(g-i)}_{R}=0.91$ and $\overline{(g-i)}_{VR}=1.42$.}\label{colordist}
\end{figure}

The distribution of colors $g-i$ for all the hot KBOs detected by our survey is shown in Figure~\ref{colordist} with the unfilled histogram. The bimodality of the distribution is clearly discernible, with two color modes centered at around $g-i=0.9$ and $g-i=1.3$. Moreover, the bimodality is present throughout the entire magnitude range covered by our survey  ($H=6.5-10.4$). This color distribution is indicative of two subpopulations within the small hot KBO population whose members have characteristically different surface colors. Hereafter, we refer to these subpopulations as the red (R) and very red (VR) KBOs. 

To carry out a more detailed analysis of the individual subpopulations, we first filter our survey KBO dataset by removing the low signal-to-noise objects. The photometric errors of the detected objects become large with decreasing apparent brightness; the color uncertainties are also increased for objects located in images that were obtained when the seeing was poor. A sizeable fraction of the objects in the survey KBO dataset have color uncertainties that are comparable to the difference between the two color modes. These large uncertainties contribute additional scatter to the color distribution and prevent clear categorization of those objects as R or VR. For the results presented below, we constructed a filtered KBO dataset by only considering the 118 objects with $\sigma_{g-i}\le 0.15$.

The color distribution of objects in the filtered KBO dataset is shown by the solid histogram in Figure~\ref{colordist}. To quantitively determine the mean colors of the R and VR subpopulations, we followed the technique employed in our past studies of bimodal asteroid populations \citep{wong,wong2} and fit the filtered KBO color distribution to a two-peaked Gaussian probability function. A standard log-probability minimization yielded the mean R and VR colors: $\overline{(g-i)}_{R}=0.91$ and $\overline{(g-i)}_{VR}=1.42$. The standard deviation of the two color modes are $\sigma_{g-i,R}=0.12$ and $\sigma_{g-i,VR}=0.09$, respectively.

This normal mixture model fitting also allows us to determine the significance of the bimodality. We computed the Bayesian Information Criterion (BIC) for both unimodal and bimodal distribution fits to the KBO color distribution. The BIC is defined as BIC $\equiv-2\log(L)+k\log(n)$, where $L$ is the likelihood of the best-fit solution, $k$ is the number of free parameters, and $n$ is the number of data points. The difference in BIC between the unimodal and bimodal fits is $\Delta$BIC$=23.0$, which shows that the bimodal interpretation is very strongly favored ($\Delta$BIC$>5$ indicates a strong preference for the model with smaller BIC).

Having established the mean colors, we can now sort individual objects into the R and VR subpopulations based on their measured colors. We classified all objects in the filtered KBO dataset with $g-i\le \overline{(g-i)}_{R}+\sigma_{g-i,R}=1.03$ as R and all objects with $g-i\ge \overline{(g-i)}_{VR}-\sigma_{g-i,VR}=1.31$ as VR. While our previous studies used the mean colors as the cutoff locations for classifying objects into subpopulations, limiting the filtered KBO dataset to just objects with $\sigma_{g-i}\le 0.15$ enabled us to extend our categorization into the region between the two mean colors. As a result, we categorized 63 objects as R and 30 objects as VR.

\subsection{Magnitude distributions}\label{subsec:mags}

Our moving object detection procedure returned the right ascension and declination of all objects in each of the four exposures taken at the corresponding fields. From here, we converted to geocentric ecliptic coordinates to obtain the ecliptic longitude $\lambda$ and latitude $\beta$ of every object. Having computed the heliocentric distance $d$ through our orbit fitting routine (Section~\ref{sec:obs}), we calculated the geocentric distance $\Delta$ of each object using standard coordinate transformations:
\begin{align}\label{geocentric}\Delta = r_{S}\cos(\beta)&\cos(\lambda-\lambda_{S}) \notag \\
&+\sqrt{r_{S}^{2}\left\lbrack\cos^{2}(\beta)\cos^{2}(\lambda-\lambda_{S})-1\right\rbrack+d^{2}},\end{align}
where $r_{S}$ and $\lambda_{S}$ are the geocentric distance and geocentric ecliptic longitude of the Sun at the time of each observation. We first interpolated the fluxes in $g$-band and $i$-band to derive the $V$-band apparent magnitude, $v$, for each object, following the spectral slope computation described in Section~\ref{subsec:centaurs}. Next, we converted the apparent $v$ magnitudes to absolute $H$ magnitudes via 
\begin{equation}\label{conversion}H=v-2.5\log_{10}(d^{2}\Delta^{2}).\end{equation}
For the identified KBOs in our dataset with previously obtained visible-band or multi-band photometry, our calculated absolute magnitudes agree the Minor Planet Center magnitudes to within 0.3~mag.

\begin{figure}[t!]
\begin{center}
\includegraphics[width=9cm]{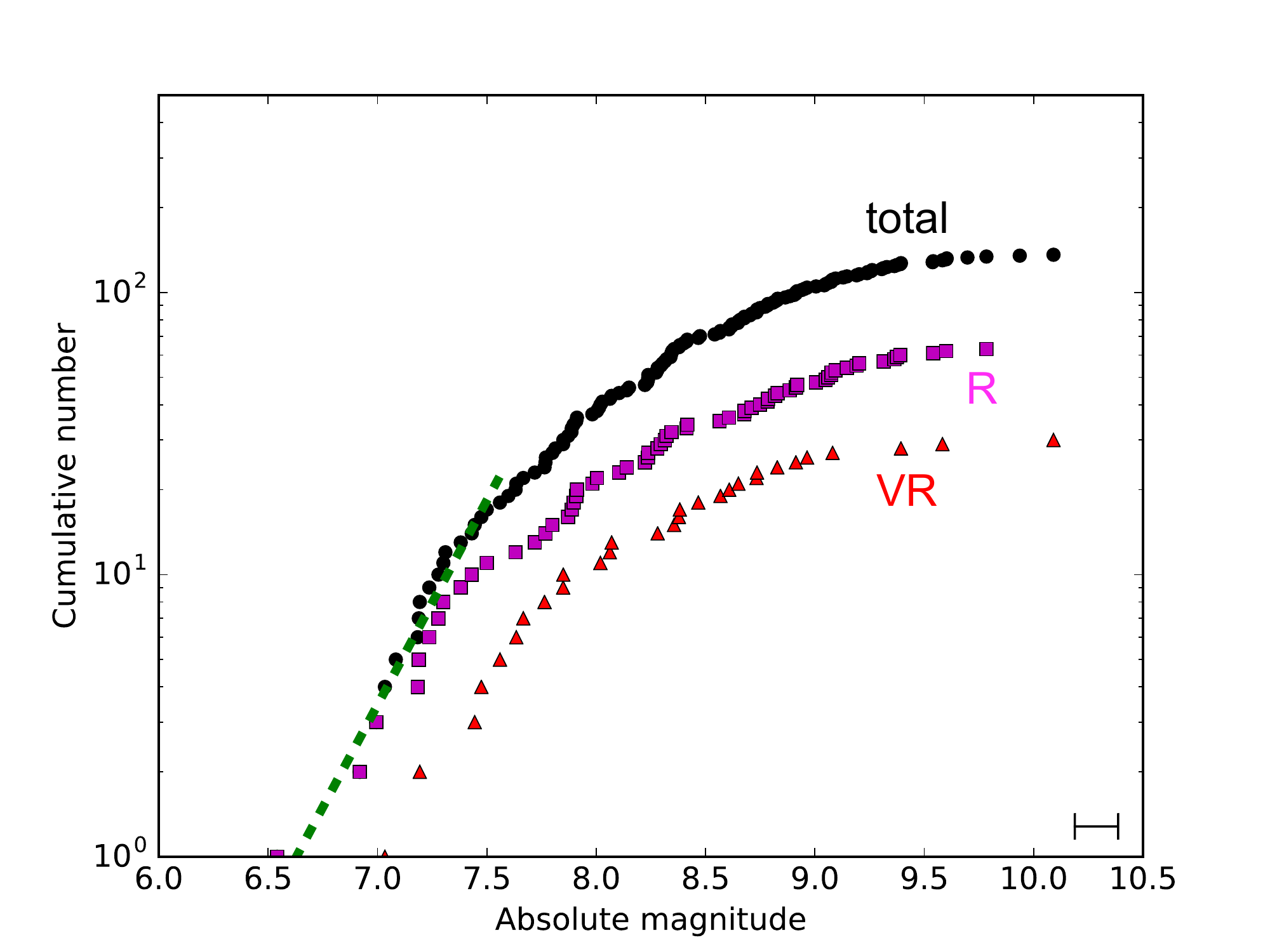}
\end{center}
\caption{Absolute $H$ magnitude distributions of the total survey hot KBO dataset (black points), as well as the categorized R and VR subpopulations (magenta squares and red triangles, respectively). The error bar in the lower right represents the typical magnitude uncertainty: 0.2~mag. The two color-magnitude distributions have very similar shapes throughout the entire magnitude range. The dashed green line indicates the best-fit power law distribution to the total magnitude distribution through $H_{\mathrm{max}}=7.5$, which has a power law slope of $\alpha=1.45$.}\label{magdist}
\end{figure}

Figure~\ref{magdist} shows the cumulative absolute magnitude distribution $N(H)$ of all objects in our survey KBO dataset (black points). The gentle rollover starting at $H\sim 8.0$ indicates the onset of detection incompleteness in our survey. The magnitude distribution of KBOs has been the subject of intensive study over the last decade, with numerous dedicated surveys having been undertaken to explore the shape of the distribution through $H = 11-12$. Many recent published works in the literature present incompleteness-corrected KBO magnitude distributions extending well beyond the completeness limit of our photometric survey \citep[e.g.,][]{fraser,shankman}.

For the sake of comparison with previous studies of the hot KBO population, we chose a conservative threshold magnitude of $H_{\mathrm{max}}=7.5$ and fit the differential magnitude distribution, $\Sigma(H)=dN(H)/dH$, of our survey KBOs through $H_{\mathrm{max}}$ with a single power law distribution of the form $\Sigma(\alpha,H_{0}|H)=10^{\alpha(H-H_{0})}$, where $\alpha$ is the power law slope, and $H_{0}$ is the normalization magnitude for which $\Sigma(H_{0})=1$. Following the methods described in \citet{wong2}, we used a Markov Chain Monte Carlo algorithm to estimate the best-fit parameters and corresponding $1\sigma$ uncertainties, which yielded $\alpha=1.45^{+0.40}_{-0.37}$. This value is steeper than the slope of $0.87^{+0.07}_{-0.2}$ reported by \citet{fraser} for the same magnitude range, though still consistent at the $1.5\sigma$ level. 

Our analysis of the $g-i$ color distribution and the classification of individual hot KBOs as R or VR allow us carry out the first study of the individual magnitude distributions of the R and VR color subpopulations. The cumulative absolute magnitude distributions of the categorized R and VR KBOs are shown in Figure~\ref{magdist}. It is immediately apparent that the R and VR magnitude distributions have nearly identical shapes throughout the enter magnitude range covered by our survey. The similarity between the R and VR magnitude distributions has important implications regarding the origin of the color bimodality, which we discuss in the next section.

Although not all KBOs in our filtered sample were classified as R or VR due to their intermediate measured colors, this incompleteness in categorization is magnitude-independent and therefore has no effect on the relative shapes of the color-magnitude distributions. Meanwhile, non-uniform seeing and differing $g$- and $i$-band signal-to-noise can cause survey biases. Among the objects in our filtered sample, the majority had the worst seeing during $g$-band observations (see supplemental table). Overall, when the seeing during $g$-band and $i$-band observations were comparable, the signal-to-noise is slightly lower in $g$-band. Both of these trends establish $g$ as the limiting band for detection. For a KBO with a given absolute magnitude near the detection limit, R objects are roughly 0.1~mag brighter in $g$-band than VR objects. 

Thus, we conclude that our survey has a bias toward R objects. We note, however, that this bias only affects the faintest objects in our sample. Therefore, the relative shape of the R and VR magnitude distributions and the R-to-VR number ratio among the brightest objects should not be influenced by the bias.

We used mean R and VR geometric albedos from the literature \citep[$p_{v,R}= 0.06$ and $p_{v,VR}=0.12$, respectively, as computed in][]{fraser} to convert absolute magnitudes to approximate diameters via the relation $D=1329\times 10^{-H/5}/\sqrt{p_{v}}$~km.
To obtain our first estimate of the R-to-VR number ratio $r$, we considered only categorized KBOs larger than 100~km in diameter, corresponding to threshold absolute magnitudes of $H=8.7$ and $H=7.9$ for R and VR objects, respectively. In this range, we detected 36 R and 10 VR objects, corresponding to $r=3.6$. Since our method for classifying of objects into subpopulations relied on the results of a double Gaussian fit (see Section~\ref{subsec:colors}), the number ratio we calculate can be affected by deviations in the measured color distribution from a true two-peaked Gaussian shape (e.g., slight asymmetries relative to mean R and VR colors). 

As a second estimate of the R-to-VR number ratio, we calculated the error-weighted mean color of all 46 objects in our filtered KBO sample larger than 100~km --- $\overline{g-i} = 1.04$. We modeled the mean color as a mixture of R and VR objects with typical colors $\overline{(g-i)}_{R}=0.91$ and $\overline{(g-i)}_{VR}=1.42$, respectively: $\overline{g-i}=q\times\overline{(g-i)}_{R}+(1-q)\times\overline{(g-i)}_{VR}$, where $q$ is the R object fraction and is related to $r$ via $r=q/(1-q)$. Through this procedure, we obtained $r=3.0$. Acknowledging the sensitivity of the computed number ratio to the particular method of calculation, we give the general conclusion that the R hot KBO subpopulation is roughly three to four times the size of the VR subpopulation.

\section{Discussion}\label{sec:discussion}
Our two-day HSC survey represents the most systematic photometric survey of small KBOs to date, greatly increasing the number of measured KBO colors for objects fainter than $H\sim 7$. We demonstrated incontrovertibly for the first time the robust bimodality in optical color among small KBOs in the hot population. The classification of objects into the R and VR subpopulations also enabled the first look at the individual color-magnitude distributions. Building on the analysis presented in the previous section, we are now in a position to compare the small hot KBOs to other populations and discuss the origin of the attested color bimodality.

\subsection{Comparison with Centaurs}\label{subsec:centaurs}

Constraining for relatively high signal-to-noise objects via the condition $\sigma_{g-i}\le 0.15$, we obtained a filtered KBO set of 118 objects (Section~\ref{subsec:colors}). Applying the same constraint to the 16 Centaurs detected by our survey yields 15 objects. 

In order to combine our data with previously published color measurements of Centaurs and hot KBOs compiled in the literature, we must convert our $g-i$ colors to spectral slopes $S$. We followed the prescription described in detail in \citet{wong} and first converted the measured apparent magnitudes to flux, normalizing to unity in $g$ band: $F_{g}=1$, $F_{i}=10^{0.4((g-i)-0.55)}$, where we have removed the solar $g-i$ color \citep{ivezic}. The associated relative flux uncertainties were computed using the second-order method of \citet{roig}: $\Delta F_{g}/F_{g}=\sqrt{2}\sigma_{g}$, $\Delta F_{i}/F_{i}=0.9210\sigma_{g-i}(1+0.4605\sigma_{g-i})$. We then calculated the spectral slope $S$ using standard linear regression.

\begin{figure}[t!]
\begin{center}
\includegraphics[width=9cm]{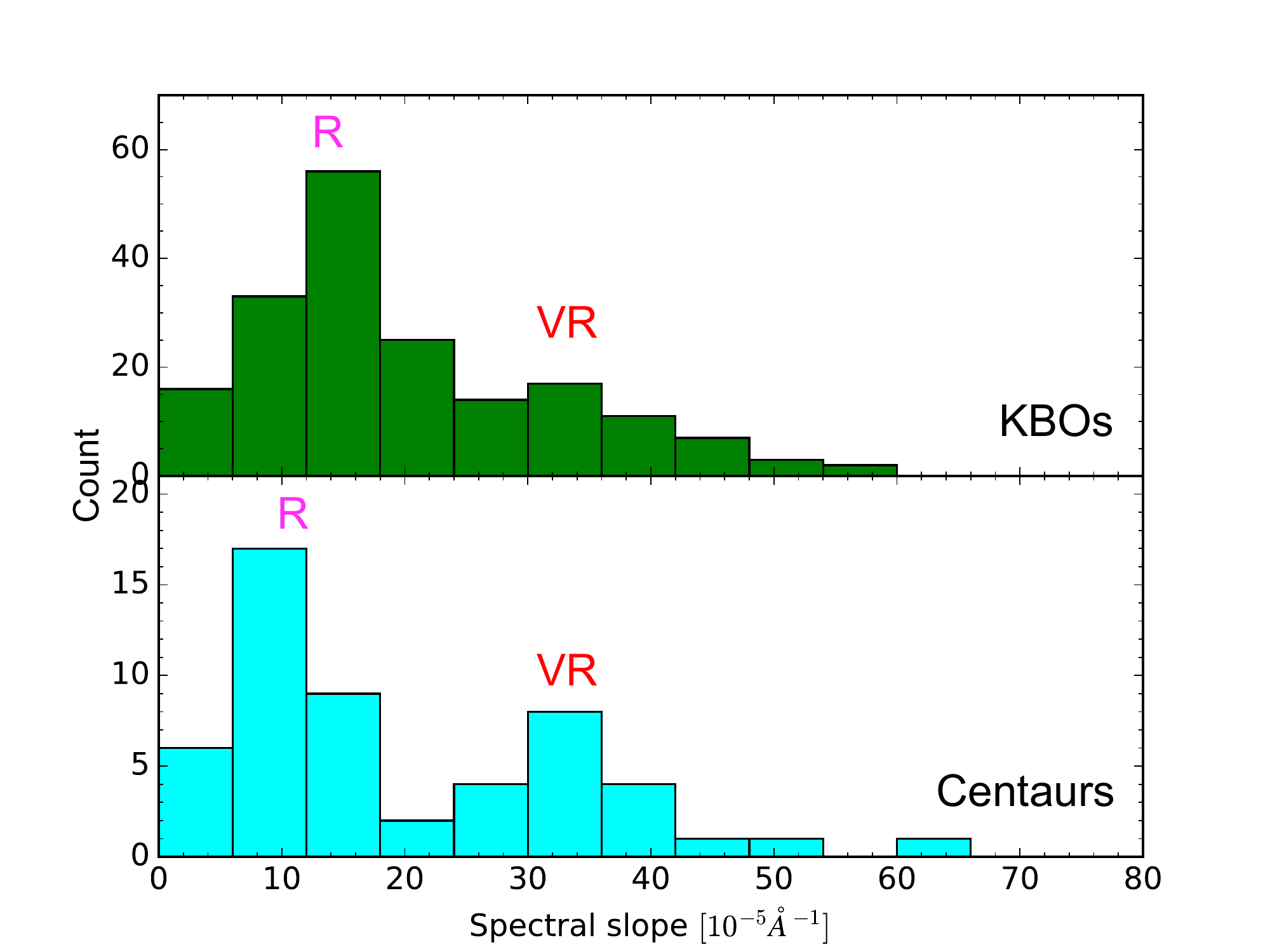}
\end{center}
\caption{Top panel: spectral slope distribution of 200 hot KBOs, incorporating objects in the filtered KBO dataset from our survey as well as objects contained in the \citet{peixinho2015} database spanning the same magnitude range. Bottom panel: same as top panel, but for 53 Centaurs detected by our survey or listed in the \citet{peixinho2015} database (with $H\ge 7$). All of these spectral slope values have uncertainties less than $10\times 10^{-5}~\AA^{-1}$. Both spectral slope distributions are bimodal, with the R and VR modes in each population located at similar spectral slope values.}\label{centaurcomparison}
\end{figure}

We queried the database of objects analyzed in \citet{peixinho2015} for hot KBOs (inclinations greater than or equal to $5^{\circ}$; Section~\ref{sec:obs}) and Centaurs, both with the condition $H \ge 7$, finding 82 and 38 objects, respectively. Combining these objects with our survey data, we constructed a total spectral slope dataset for 200 hot KBOs and 53 Centaurs. The corresponding spectral slope distributions are shown in Figure~\ref{centaurcomparison}.

We note that the transformation from $g-i$ color to $S$ is not linear, with the distance between spectral slope values corresponding to evenly spaced $g-i$ color values being amplified as the color becomes redder. As a result, the shape of the VR mode in the spectral slope distribution is broad and noticeably asymmetric, as opposed to the shape of the analogous peak in the $g-i$ color distribution (Figure~\ref{colordist}). This serves to reduce the saliency of the bimodality in the spectral slope distribution. However, even with this effect, the bimodality in spectral slope among the hot KBOs is still discernible.

The bimodality in spectral slope among the Centaurs is well-attested by numerous previous studies \citep[e.g.,][]{peixinho2,barucci3,perna}. The comparison of Centaurs with hot KBOs demonstrates that not only is the hot KBO spectral slope distribution also bimodal, the two color modes closely correspond in location to the two color modes in the Centaur spectral slope distribution. In other words, the R and VR Centaurs (often referred to as gray/neutral and red in the literature) have the same colors as R and VR hot KBOs. 

From this observation, we can conclusively resolve the issue of the color bimodality of Centaurs in this magnitude range (which includes all known Centaurs except Chariklo and Chiron) ---- before being scattered inward onto their current comet-like orbits, R and VR Centaurs were members of the R and VR hot KBO subpopulations, respectively, as previously suggested by \citet{fraserbrown} and \citet{peixinho}.

\subsection{Comparison with larger hot KBOs}\label{subsec:largehot}
\begin{figure}[t!]
\begin{center}
\includegraphics[width=9cm]{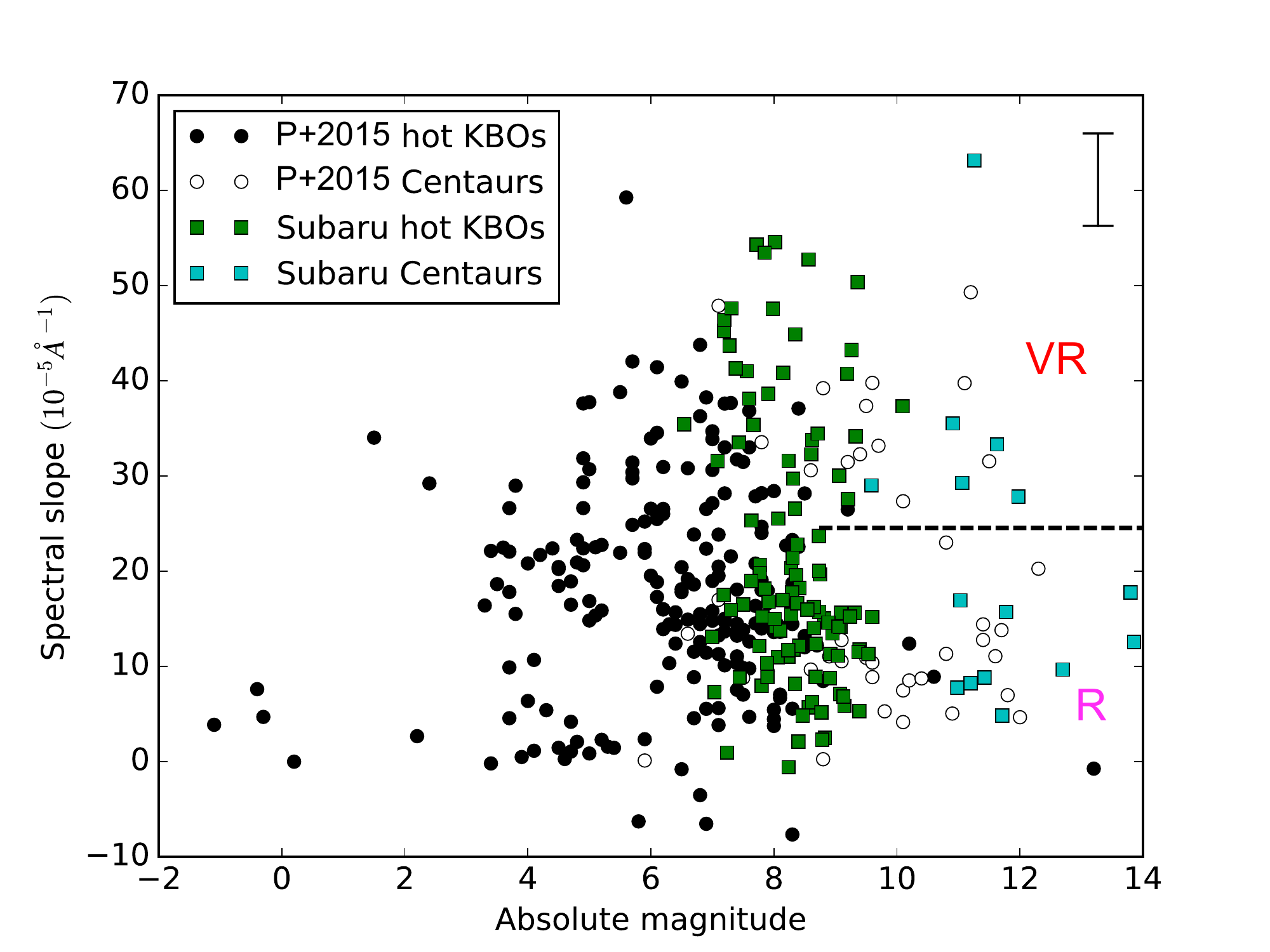}
\end{center}
\caption{Color-magnitude plot for hot KBOs and Centaurs combining data tabulated in \citet{peixinho2015} (dots) and the results of our Subaru survey (squares). Only spectral slope measurements with uncertainties less than $10\times 10^{-5}~\AA^{-1}$ are included; error bars on data points are omitted for the sake of clarity, with the size of the maximum uncertainty indicated by the example error bar in the upper right. Bimodality in color is discernible among all objects fainter than $H\sim 7$. The horizontal dashed line indicates the approximate boundary between R and VR colors. Meanwhile, the larger hot KBOs display a uniform color distribution, with the exception of a small clustering of neutral colored objects in the range $H=2-6$ that is dominated by Haumea collisional family members.}\label{hotKBOcomparison}
\end{figure}

We also place the color data of small KBOs from our survey in the context of the wider hot KBO population as a whole. Figure~\ref{hotKBOcomparison} shows the color-magnitude distribution of all hot KBO and Centaurs with spectral slope errors less than $10\times 10^{-5}~\AA^{-1}$. The plot illustrates how our survey data extends the known body of high quality KBO color measurements into the size region that had previously only been adequately studied using Centaur colors as proxy.

For both hot KBOs and Centaurs, a bimodality in color is evident at all magnitudes among objects fainter than $H\sim 7$, as was discussed in Section~\ref{subsec:centaurs}. The Subaru hot KBO data nearly triple the number of high signal-to-noise KBO color measurements between $H=7$ and $H=9$ contained in the \citet{peixinho2015} database and are instrumental in solidifying the observation of bimodality in that size range, which was previously hinted at using far fewer data points \citep{fraserbrown,peixinho,peixinho2015}. 

Notably, however, the colors of larger, brighter hot KBOs show a more uniform distribution, with many objects having spectral slope values intermediate between the two color types. The apparent clustering of near-neutral objects between $H=2$ and $H=6$ is primarily due to the presence of the Haumea collisional family. Previous studies have shown that, omitting the Haumea family members, no significant bimodality is present among objects brighter than $H\sim 6$ \citep{fraserbrown,peixinho,peixinho2015}. In other words, only small hot KBOs are bimodal in color.

The threshold of $H\sim 6$ corresponds to a diameter of roughly $300-400$~km. Observations of these large- and intermediate-sized objects have revealed somewhat higher albedos than their smaller counterparts \citep{stansberry}. This suggests that the larger KBOs have fundamentally different surface properties than the objects covered by our survey. The discrepancy may be due to the effects of secondary processes since formation, such as the onset of cryovolcanism or differentiation, which would not have occurred in smaller bodies \citep[see, for example, discussion in][]{fraserbrown}.

\subsection{Comparison with Jupiter Trojans}\label{subsec:trojans}

Recent theories of solar system evolution posit that a mean motion resonance crossing between Jupiter and Saturn roughly 600~Myr after the formation of the solar system led to significant alterations in the orbital architecture of the giant planets \citep{morbidelli,gomes,tsiganis}. An important consequence of this dynamical restructuring is the scattering of the primordial body of planetesimals that were situated beyond the orbits of the ice giants. While the majority of the mass within this region was expelled from the solar system as the orbits of the ice giants expanded, a fraction of the planetesimals were scattered outward to become the present-day Kuiper Belt. The remainder was scattered inward and eventually captured into resonance with Jupiter to become the Trojan asteroids. Therefore, the current paradigm of solar system evolution predicts a common origin for Jupiter Trojans and KBOs.

Several authors have previously noted that the overall magnitude distributions of Trojans and hot KBOs brighter than $H\sim 8-9$ are similar \citep[e.g.,][]{fraser}. In \citet{wong}, we reported a power law slope of $1.11\pm 0.20$\footnotemark[1] for Trojans brighter than $H\sim 8$. In our present study, we obtained a best-fit magnitude distribution slope of $1.45^{+0.40}_{-0.37}$ for hot KBOs brighter than $H = 7.5$. The comparison demonstrates that the bright hot KBOs and Trojans of the same size have statistically indistinguishable magnitude distributions.

\footnotetext[1]{In \citet{wong}, the uncertainty in the slope was reported erroneously as 0.02, instead of the correct value 0.20.}

Similar to the KBOs, the Trojans display a robust optical color bimodality, albeit with the peaks of the modes centered at bluer colors \citep[$g-i=$ 0.73 and 0.86;][]{wong2}. In \citet{wong}, we categorized Trojans into the two respective color subpopulations and analyzed the individual color-magnitude distributions. We showed that the two Trojan subpopulations have statistically indistinguishable magnitude distributions in the bright end ($H< 9$), just as the R and VR hot KBO subpopulations have nearly identical magnitude distribution shapes throughout the brightness range covered by our survey (Section~\ref{subsec:mags}). 

Taken together, the analyses of Trojans and KBOs provide strong evidence to support one of the key predictions of current models of solar system evolution. Not only are the overall magnitude distributions of the hot KBOs and Trojans similar, but the individual constituent color subpopulations have comparable magnitude distributions. Therefore, we can posit that each of the two subpopulations in the Trojans and hot KBOs are sourced from a single progenitor body of planetesimals. 

\subsection{Origin of the color bimodality}

The bimodal color distribution of hot KBOs is indicative of two different surface types, with distinct photometric properties. There are two classes of possible explanations for the origin of the observed color bimodality: (1) the R and VR KBOs formed in different regions of the protoplanetary disk and were sourced from material with different bulk compositions, and (2) the R and VR KBOs formed in the same region of the disk and out of the same material, but later developed distinct surface properties via some secondary evolutionary process.

The size distribution, or by proxy, the magnitude distribution, of a population contains information about both the accretion environment and the subsequent collisional evolution of the population. Therefore, the color-magnitude distributions we derived in our analysis are a crucial tool in helping us understand how the color bimodality developed. 

In Figure~\ref{magdist}, we showed that the R and VR hot KBO magnitude distributions are nearly identical in shape, which suggests that the two subpopulations have undergone a similar formation and evolutionary history and therefore likely formed out of the same body of material within the protoplanetary disk. This similarity in magnitude distribution shape contrasts with, for example, the strong discrepancy between the hot and cold KBO magnitude distributions \citep[e.g.,][]{fraser}. 

In \citet{wong3}, we proposed a simple hypothesis, that accounts for the development of a color bimodality within the primordial planetesimal disk beyond the orbits of the ice giants. Assuming the dynamical framework of the Nice Model, the model describes a primordial body of material situated between 15 and roughly 30~au, from which the modern hot KBO population is sourced \citep[e.g.,][]{morbidelli,levison2008}. After a residence time of roughly 600~Myr \citep{gomes}, a period of chaotic dynamical evolution throughout the outer solar system scattered a fraction of these objects into the current Kuiper belt region (see discussion in previous subsection).

The hypothesis posits that all the bodies within the primordial trans-Neptunian disk accreted with a roughly cometary composition, i.e., rocky material and water ice, with a significant volume of other volatile ices such as carbon dioxide, methanol, ammonia, hydrogen sulfide, etc. During the accretion process, large objects migrated differentially through the disk \citep[e.g.,][]{kenyon}, so that the overall disk was well-mixed compositionally, with a single characteristic size distribution. Once these bodies were exposed to insolation after the dispersal of the gas disk, sublimation-driven loss of volatiles from the outermost layers led to the development of distinct surface compositions throughout this region. 

The loss rate of each individual volatile ice species is correlated with its vapor pressure, which in turn has a strong exponential dependance on temperature. As a result, each volatile ice species would have a sublimation line at a different location within the primordial disk: objects situated inward of the line became depleted in that species on their surfaces, while objects situated beyond the line retained that species on their surfaces. 

From our modeling, we found that the sublimation line of hydrogen sulfide ice (H$_{2}$S) passed through the region of the primordial trans-Neptunian planetesimal disk. Therefore, objects within this single primordial population would have been divided into two groups --- those that retained H$_{2}$S on their surfaces, and those that did not. Previous experimental work has shown that irradiation of volatile ices leads to a general redding of the optical spectral slope \citep[e.g.,][]{brunetto}. We posited that the presence of H$_{2}$S on the surface of an object would contribute additional reddening relative to case where H$_{2}$S was absent. H$_{2}$S is known to induce a strong reddening of the surface upon irradiation, as has been observed in the H$_{2}$S frost deposits on Io \citep{carlson}.

After the Nice Model dynamical instability, some objects in the trans-Neptunian region were scattered outward, sharing a common size distribution by virtue of their common formation environment while inheriting the surface color bimodality that developed in the primordial planetesimal disk due to differential loss/retention of H$_{2}$S. Likewise, some objects were scattered inward to eventually settle into resonance with Jupiter, thereby explaining the observed color bimodality in the Trojans and the similarity in magnitude distributions between the color subpopulations of the Trojans and hot KBOs. In \citet{wong3}, we speculate that the bluer colors of the two Trojan subpopulations may be due to secondary processing of the irradiated surfaces following their emplacement into a higher-temperature environment at 5.2~au.

\section{Conclusion}
In this paper, we presented the results of a wide-field photometric survey of KBOs carried out using the Subaru/HSC instrument. Over the course of two nights of observation, we imaged 356 KBOs in the $g$ and $i$ filters, from which we obtained the $g-i$ colors. After constraining our dataset to the hot KBO population (objects with inclination greater than or equal to $5^{\circ}$) and filtering out low signal-to-noise detections, we examined the resulting color distribution of 118 objects and provided the first separate analysis of the magnitude distributions of the two attested subpopulations within the hot KBO population --- the R and VR subpopulations. The main conclusions are summarized below:
\begin{itemize}
\item The optical color distribution of hot KBOs shows a robust bimodality across the entire magnitude range covered by our survey ($H>6.5$). The two modes in the color distribution correspond to the R and the VR subpopulations and have mean colors of $\overline{(g-i)}_{R}=0.91$ and $\overline{(g-i)}_{VR}=1.42$, respectively.
\item The absolute magnitude distributions of the R and VR subpopulations are nearly identical in shape throughout the magnitude range covered by our survey. The estimated R-to-VR number ratio is between 3-to-1 and 4-to-1.
\item The color distributions of Centaurs and similarly-sized hot KBOs are comparable in shape. The locations of the R and VR color modes are similar for both populations and indicate that the R and VR Centaurs are sourced from the R and VR hot KBO subpopulations, respectively.
\item Unlike their smaller counterparts, hot KBOs larger than $H\sim 6$ have a more uniform color distribution when excluding the neutral-colored members of the Haumea collisional family.
\item The magnitude distributions of the R and VR hot KBOs are similar to the magnitude distributions of the two color subpopulations attested in the Jupiter Trojans, suggesting a common origin for the individual color subpopulations, as predicted by recent theories of solar system evolution.
\item The similarity between the R and VR KBO magnitude distributions suggests that the two subpopulations share a single progenitor population and have undergone similar formation and evolutionary histories. We hypothesize that the color bimodality arose due to location-dependent retention vs. loss of H$_{2}$S on the surfaces of these bodies in the early solar system, prior to the onset of dynamical instability that emplaced them into their current location.
\end{itemize}

\section*{Acknowledgments}
This research was supported by Grant NNX09AB49G from
the NASA Planetary Astronomy Program and by the Keck Institute for Space Studies.

\end{document}